\documentclass[10pt,3p,numbers]{elsarticle} 
\usepackage[T1]{fontenc}
\usepackage[utf8]{inputenc}
\usepackage{newtxtext,newtxmath}

\usepackage[final,protrusion=true,expansion=true,tracking=true,kerning=true,spacing=true]{microtype}
\usepackage{amssymb}
\usepackage{amsmath, amsfonts, amsthm}
\usepackage{graphicx}
\usepackage[dvipsnames]{xcolor}
\usepackage{booktabs}
\usepackage[ruled,vlined,linesnumbered]{algorithm2e} 

\SetNlSty{text}{}{}

\SetKwInput{KwInput}{Input}
\SetKwInput{KwOutput}{Output}
 
\DeclareMathOperator{\bigO}{\mathcal{O}}

\usepackage{booktabs}
\usepackage{tabularx}

\usepackage{float}
\usepackage{caption}
\usepackage[dvipsnames]{xcolor} 
\usepackage[pdfencoding=auto,psdextra,unicode]{hyperref}

\usepackage[capitalise]{cleveref}
\crefname{section}{Sec.}{Secs.}
\crefname{subsection}{Sec.}{Secs.}
\crefname{subsubsection}{Sec.}{Secs.}
\crefname{equation}{Eq.}{Eqs.}
\crefname{figure}{Fig.}{Figs.}
\crefname{table}{Table}{Tables}
\crefname{appendix}{SI}{SIs.}

\hypersetup{colorlinks=true, linkcolor=MidnightBlue,citecolor=MidnightBlue, urlcolor=MidnightBlue}

\usepackage{lineno}
\usepackage{etoolbox}

\journal{Journal Name} 

\makeatletter
\def\ps@pprintTitle{%
 \let\@oddhead\@empty
 \let\@evenhead\@empty
 
 \let\@evenfoot\@oddfoot}
\makeatother

\makeatletter
\let\@afterindenttrue\@afterindentfalse
\makeatother

\begin{document}

\begin{frontmatter}

 \title{Extreme Event Precursor Prediction in Turbulent Dynamical Systems via CNN-Augmented Recurrence Analysis}
\author[inst1]{R. Agarwal}
\author[inst1]{M.A. Mohamad\corref{cor1}}

\cortext[cor1]{Corresponding author. Email: \href{mailto:mustafa.mohamad@ucalgary.ca}{mustafa.mohamad@ucalgary.ca}}

\affiliation[inst1]{
    organization={Department of Mechanical and Manufacturing Engineering, University of Calgary}, 
    city={Calgary}, 
    state={AB}, 
    country={Canada}
}

\begin{abstract}
\noindent
We present a general framework to predict precursors to extreme events in turbulent dynamical systems.  The approach combines phase-space reconstruction techniques with recurrence matrices and convolutional neural networks to identify precursors to extreme events. We evaluate the framework across three distinct testbed systems: a triad turbulent interaction model, a prototype stochastic anisotropic turbulent flow, and the Kolmogorov flow. This method offers three key advantages: (1) a threshold-free classification strategy that eliminates subjective parameter tuning, (2) efficient training using  only $\bigO(100)$ recurrence matrices, and (3) ability to generalize to unseen systems. The results demonstrate robust predictive performance across all test systems: 96\% detection rate for the triad model with a mean lead time of 1.8 time units, 96\% for the anisotropic turbulent flow with a mean lead time of 6.1 time units, and 93\% for the Kolmogorov flow with a mean lead time of 22.7 units.

\end{abstract}


\end{frontmatter}

\section{Introduction}
Turbulence is ubiquitous in natural and engineering systems, appearing across a range of scales, from small-scale vortical structures in stirred fluids to the large-scale dynamics of ocean-atmospheric systems. The nonlinear dynamics of turbulent flows exhibits a hierarchy of instabilities,  including both persistent phenomena (associated with energy cascades) and intermittent episodes that significantly alter flow characteristics. Such intermittent events arise in a variety of contexts, from large-scale atmospheric blocking patterns~\cite{davini2012} to localized dissipation bursts in near-wall turbulent flows~\cite{jimenez2001}.

Extreme events in fluid flows, water waves, and engineering systems that interact with these media constitute a topic of vital importance, not only for risk evaluation, optimization, and design, but also for predictive control and autonomy. These systems exhibit a spectrum of extreme phenomena that span various temporal and spatial scales. Examples include extreme dissipation and enstrophy events in hydrodynamic turbulence~\cite{ishihara2007}; cavitation inception arising from low pressures in turbulent shear layers~\cite{iyer2002}; high-energy acoustic bursts from turbulent jets~\cite{schmidt2019}; transitions between chaotic and regular regimes in vortex-induced vibrations (VIV) with implications for structural fatigue~\cite{modarres-sadeghi2010}; extreme ship motions and loads in irregular wave fields~\cite{belenky2019,mohamad2016}; and atmospheric blocking events linked to extreme weather phenomena~\cite{davini2012}.

These extreme events are typically governed by continuous dynamical systems described by partial differential equations that evolve in an infinite-dimensional function space. In geophysical settings and high Reynolds number engineering flows, the number of active degrees of freedom becomes exceptionally large due to broadband nonlinear interactions across scales. Consequently, the application of dimensional reduction techniques that preserve extreme events presents significant challenges; attempts to represent such systems with low-dimensional models may reproduce typical (average) behavior, but often fail to capture the rare and intermittent instabilities that underlie extreme events~\cite{sapsis2013}.

We characterize extreme events as large excursions of a system observable, $q(u)$, where $u \in \mathbb{R}^N$ is the full state vector of the system.  These excursions are typically rare and intermittent, that deviate from the mean by multiple standard deviations. The functional form of $q(u)$ depends on application-specific requirements and typically results in a nonlinear expression. Different physical contexts necessitate different observables: for example, researchers studying rogue waves may be interested in the free surface elevation, while studies of fluid–structure interactions might examine the force distributions on structural elements~\cite{mohamad2016a,mohamad2018}.

Current predictive strategies for extreme events have typically relied on statistical or sampling based methods. Key among these are extreme value theory (EVT) and large deviation theory (LDT); these frameworks characterize rare event probabilities through the computation of initial conditions most likely to transition towards extreme states~\cite{doan2021}. Such statistical approaches have successfully identified precursors for phenomena including turbulent channel flow relaminarization~\cite{blonigan2019} and nonlinear rogue waves~\cite{dematteis2018}. In parallel, data-driven methodologies have emerged as an alternative approach. Machine learning, in particular, has shown promise in predicting the dynamics of chaotic flows, achieving accurate short-term forecasts while preserving long-term statistical properties. For instance, Long Short-Term Memory (LSTM) networks were employed by~\citet{vlachas2018} to predict the evolution of the Kuramoto–Sivashinsky equation and a barotropic climate model, demonstrating both short-term accuracy and convergence towards invariant measures. Similar recurrent neural network (RNN) architectures have successfully simulated shear turbulence evolution, reproducing moments of velocity statistics~\cite{Srinivasan2019}. Echo State Networks (ESNs) have  been shown to capture chaotic dynamics within the predictability horizon of chaotic systems~\cite{pathak2018,pathak2018a,doan2019}, while also being able to recover ergodic averages~\cite{huhn2020}.

Despite these recent advances, the accurate prediction and mitigation of extreme events in complex systems remains a challenge. One of the main difficulties lies in their  seemingly spontaneous emergence, occurring with few, if any,  discernible precursors, which severely limits the effectiveness of observation-based forecasting.  Moreover, while data-driven methods such as with machine learning, have shown promise in capturing chaotic behavior, their extension to rare-event regimes faces significant challenges, including insufficient statistical sampling in critical regions of phase space and poor generalization across varying operating conditions.

In response, hybrid approaches have gained traction. For instance,  \citet{wan2018} combined a reduced-order model with an LSTM to predict dissipation events in Kolmogorov flow and intermittent transitions in a barotropic model, while  \citet{farazmand2017} formulated a variational solution to identify precursors leading to a constrained optimization problem,  and also large-deviation approaches that calculate instantons to determine the most probable rare events, such as in~\cite{soons2025}. Existing studies often employ prescribed perturbations to steer the system away from extreme regimes.  Computational methods, whether purely statistical or data-driven, tailored to extreme event prediction remain in their early stages, necessitating additional approaches that integrate data-driven inference with the underlying physical structure. 

\begin{figure}[htbp]
    \centering
    \includegraphics[width=0.45\textwidth]{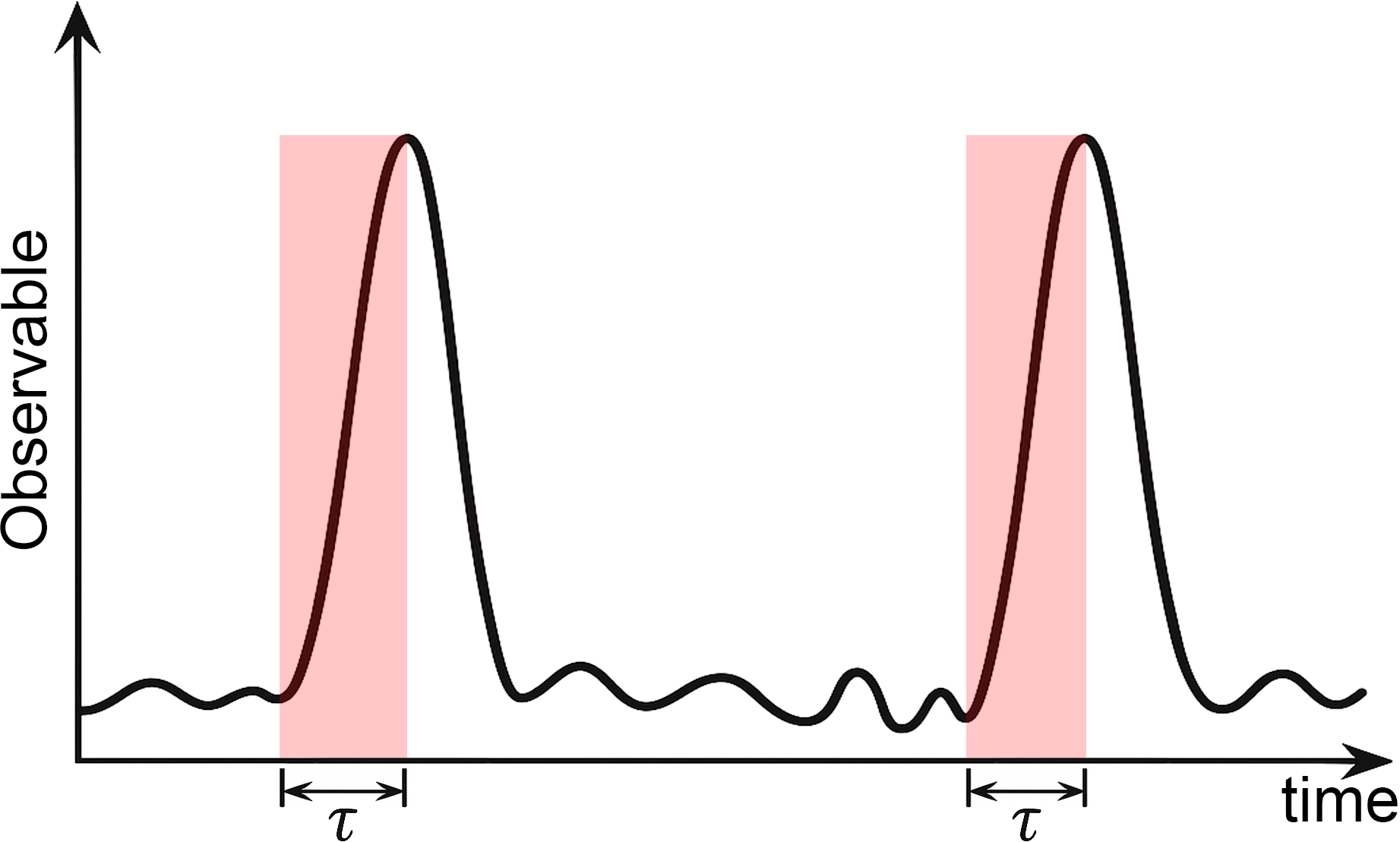}
    \caption{Illustration of intermittent bursts of an observable. The highlighted regions mark an approximation of the growth phase of the extreme events and indicate an approximate duration  $\tau$ corresponding to the growth phase.}
    \label{fig:figure_1}
\end{figure}

Here, we develop a  general  framework for predicting extreme events in turbulent dynamical systems by combining nonlinear dynamical system analysis with machine learning. As illustrated in Figure~\ref{fig:figure_1}, these events manifest as intermittent bursts in observables and correspond in state space to rapid excursions away from the background attractor. Specifically, we identify universal precursors to these extreme events through a novel combination of phase-space reconstruction, recurrence matrix analysis, and deep learning architectures. By detecting the subtle dynamical signatures that emerge during the growth phase, we can anticipate extreme events before they fully develop.
Our approach addresses several limitations of existing techniques, including reliance on arbitrary thresholds, the need for large training datasets, and poor generalization across different chaotic or turbulent systems. We demonstrate that by integrating the pattern recognition capabilities of convolutional neural networks (CNNs) with physically informed features derived from recurrence quantification analysis,  our framework can achieve accurate and computationally efficient prediction of extreme events. This methodology offers a practical tool for early warning systems in various engineering and geophysical applications where extreme events pose significant risks.

\section{Methodology}

\begin{figure}[htbp]
    \centering
    \includegraphics[width=0.85\textwidth]{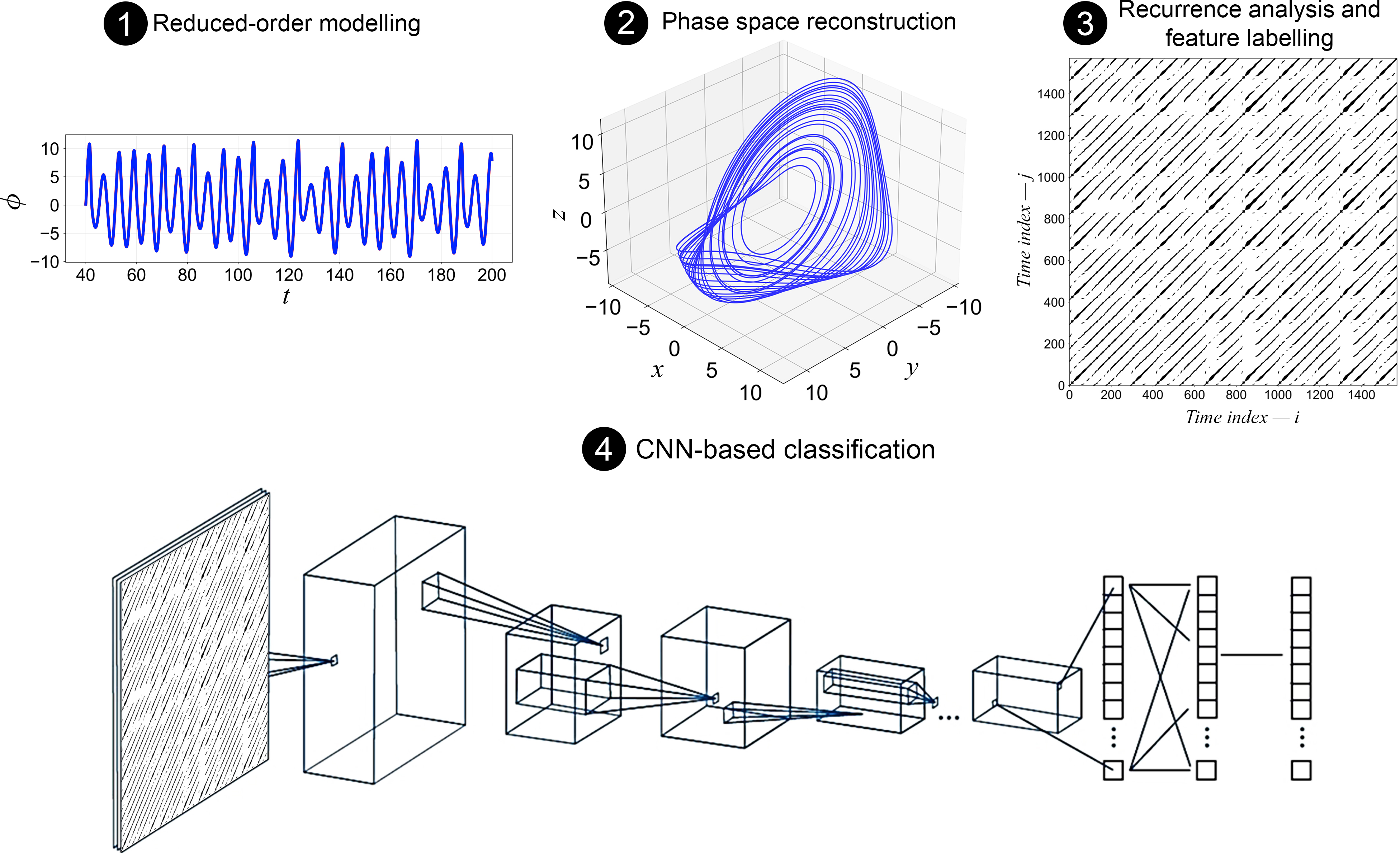}
    \caption{Schematic of the proposed framework combining nonlinear dynamics and deep learning for extreme event prediction. The approach comprises four core components: (1) reduced-order modeling with oscillator–event interactions; (2) phase space reconstruction via time-delay embedding; (3) recurrence plot analysis for detecting intermittent dynamics; and (4) CNN-based classification of precursor patterns.}
    \label{fig:figure_2}
\end{figure}

The proposed methodology integrates deep learning with nonlinear dynamical system analysis to construct a robust framework for precursor detection of extreme events. It combines the pattern recognition capabilities of convolutional neural networks (CNNs) with physically informed features extracted from recurrence quantification analysis. As illustrated in Figure~\ref{fig:figure_2}, the framework consists of four core components:

\paragraph{\normalfont\textit{Step 1: Reduced-order modeling}}

This step employs a deterministic reduced-order model that serves as a proxy for extreme event generation in more complex dynamical systems. The model captures coupled interactions between damped harmonic oscillators and discrete threshold-triggered events. A nonlinear feedback loop between continuous oscillatory modes and instantaneous impulses gives rise to rich dynamics, including self-sustained oscillations and chaos, without the need for stochastic forcing. Full details of the model are provided in~\cref{sec:section_2.1}.

\paragraph{\normalfont\textit{Step 2: Phase space reconstruction}}

In this step, we reconstruct the system's attractor in phase space via time-delay embedding techniques from time-series data, providing a geometric representation of its underlying dynamics. Optimal embedding parameters are selected using average mutual information and Cao's method. This reconstruction yields a trajectory in phase space that captures the evolution of the system. The details are present in~\cref{sec:section_2.2}.

\paragraph{\normalfont\textit{Step 3: Recurrence plot and feature labeling}}

Recurrence plot analysis is applied to the reconstructed attractor to identify patterns indicative of intermittency. We compute recurrence matrices and apply the 0-1 test for chaos and intermittency classification metrics to distinguish between dynamical regimes that precede extreme events. Details are provided in~\cref{sec:section_2.3}.

\paragraph{\normalfont\textit{Step 4: Convolutional neural network classification}}

In the final step, a CNN is trained on recurrence matrices to classify states leading to extreme events. The network extracts hierarchical spatial features from the recurrence plots and uses these to identify dynamical precursors before the events fully manifest. The implementation details are provided in~\cref{sec:section_2.4}.

\subsection{Reduced-Order Modeling}
\label{sec:section_2.1}

Many complex systems across physics, engineering, and neuroscience exhibit dynamics governed by the interplay between oscillatory modes and discrete, threshold-triggered events. This coupling gives rise to rich nonlinear behavior, including intermittency, mode-locking, and chaos. We adopt a generalized deterministic framework that captures these dynamics through interactions between damped harmonic oscillators and an event-generation mechanism. This phenomenological model is adapted from the formulation of \citet{matveev2003}.

Consider a system of oscillators with amplitudes $\eta_n(t)$ corresponding to the $n$-th mode. The dynamics of these oscillators can be described by:
\begin{equation}
\frac{d^2\eta_n}{dt^2} + 2\xi_n\omega_n\frac{d\eta_n}{dt} + \omega_n^2\eta_n = f_n(t)
\end{equation}
where $\omega_n$ is the natural frequency of the $n$-th mode, $\xi_n$ is the corresponding damping coefficient, and $f_n(t)$ is the forcing term. 

For the kicked oscillator system, the forcing term $f_n$ consists of instantaneous impulses that occur at discrete times when a system variable exceeds a critical threshold. These impulses are modeled as:
\begin{equation}
f_n(t) = c \sum_j A_j\psi_n(x_j)\delta(t - t_j)
\end{equation}
where $c$ is a coefficient, $A_j$ is the amplitude of the kick, $\psi_n(x_j)$ describes the spatial distribution of the kick on the $n$-th mode, and $\delta(t - t_j)$ denotes the Dirac delta function centered at kick time $t_j$.
Between kick events, the system evolves as a standard damped oscillator. At each kick time $t_j$, the following jump conditions apply:
\begin{align}
\eta_n(t_j^+) - \eta_n(t_j^-) &= c A_j \psi_n(x_j),\\
\dot{\eta}_n(t_j^+) - \dot{\eta}_n(t_j^-) &= 0.
\end{align}

In this simple reduced-order model for extreme event generation,   threshold crossings in the oscillator states trigger impulsive kicks, which in turn modify the state of the oscillators through instantaneous forcing. The timing and location of each kick explicitly depend on the state of the system, resulting in closed-loop interactions. This setup generates self-sustained oscillations and chaotic trajectories without requiring any stochastic input. As demonstrated in later sections, the model serves as a minimal analogue for extreme event generation in more complex systems, capturing the role of threshold-driven feedback. Further details are provided in~\cref{supp:doc1}.

\subsection{Phase-Space Reconstruction of System Dynamics}
\label{sec:section_2.2}

The dynamical behavior of a system under varying operating conditions can be analyzed by reconstructing its phase space from time-series measurements. This process, known as delay embedding~\cite{takens1981}, transforms scalar time-series data into a set of delay vectors that approximate the system's trajectory in a reconstructed phase space, where its evolution can be geometrically visualized .
The reconstructed phase space consists of delay vectors of the form
\begin{equation}
\mathbf{p}_i(d) = [x(t_i), x(t_i + \tau), x(t_i + 2\tau), \dots, x(t_i + (d-1)\tau)],
\end{equation}
where $\tau$ is the time delay and $d$ is the embedding dimension. The elements of  $\mathbf{p}_i(d)$  are the coordinates in a $d$-dimensional phase space and represent the system's state at time $t_i$. The set of these vectors over time approximates the geometry of  system's underlying attractor.
To achieve a faithful reconstruction, appropriate values for the delay $\tau_{\text{opt}}$ and the minimum embedding dimension $d$ must be selected. Typically this is done using techniques such as average mutual information and Cao’s method.


The optimal delay $\tau_{\text{opt}}$ can be estimated as the value of $\tau$ at which the average mutual information~\cite{abarbanel1993}   between delayed coordinates attains its first local minimum. The average mutual information (AMI) of a signal is given by the expression:
\begin{equation}
I(x(t), x(t+\tau)) = \sum_{ij} P_{i,j}(x(t), x(t + \tau))\log \biggl( \frac{P_{ij} (x(t), x(t + \tau))}{P_i(x(t))P_j(x(t + \tau))} \biggr)
\end{equation}
where $P_i$ denotes the probability that $x(t)$ is in bin $i$ of the histogram constructed from $x$ and $P_{ij}$ is the probability that $x(t)$ is in bin $i$ and $x(t+\tau)$ in bin $j$.
The average mutual information quantifies the amount of information between $x(t)$ and its delayed version $x(t + \tau)$. The first local minimum of $I(\tau)$ indicates the delay at which successive components of the delay vectors are least redundant.   This choice yields an embedding that maximizes information about the system's attractor in the reconstructed phase space. 


To estimate a suitable embedding dimension $d$, we use the technique developed by \citet{cao1997}.  Cao's method is   an improved version of the False Nearest Neighbors method~\cite{abarbanel1993}, where one tracks the number of false neighbors to each point in the phase space as the embedding dimension is progressively increased. A false neighbor to a point in phase space is one that moves away from it as the embedding dimension is increased. Mathematically, once the optimum time lag has been obtained (using the AMI approach described above), we can construct a measure $a(i, d)$ of the form:
\begin{equation}
a(i, d) = \frac{\lVert \mathbf{p}_i(d+1) - \mathbf{p}_{n(i,d)}(d+1)\rVert}{\lVert\mathbf{p}_i(d) - \mathbf{p}_{n(i,d)}(d)\rVert}
\end{equation}
where $i = 1, 2,\ldots, (N-d\tau)$ and $n(i, d)$ is the index of the nearest neighboring point in phase space to the point $\mathbf{p}_i$ and $\lVert\cdot\rVert$ represents the Euclidean distance between two points. The dependency on the index $i$ is removed by taking the average $a(i, d)$ obtained at different values of $i$ as:
\begin{equation}
E(d) = \frac{1}{N - d\tau_\text{opt}} \sum_{i=1}^{N-d\tau_\text{opt}} a(i, d)
\end{equation}
Here, $E(d)$ is a function  of the dimension $d$ and the optimal time lag $\tau_\text{opt}$. The variation of $E(d)$ as the dimension is increased from  from $d$ to $d+1$ is determined by defining $E_1(d)$ as:
\begin{equation}
E_1(d) = \frac{E(d+1)}{E(d)}
\end{equation}
If $E_1(d)$ stops changing when the value of $d$ is greater than $d_0$, then $d_0+1$ is chosen as the minimum embedding dimension for the time series. 

Since real-world time series data are of  finite length, it is often difficult to distinguish a stochastic signal from a deterministic signal, solely by observing the changes to $E_1(d)$ for increasing values of $d$. Whereas $E_1(d)$ tends to saturates beyond a certain $d$ for  deterministic signals, it always increases with increasing $d$ for random signals. To clearly distinguish deterministic signals from stochastic signals, we define an auxiliary measure $E_2(d)$ from the time series $x(t)$:
\begin{equation}
E_2(d) = \frac{E^*(d+1)}{E^*(d)}
\end{equation}
where
\begin{equation}
E^*(d) = \frac{1}{N - d\tau_\text{opt}} \sum_{i=1}^{N-d\tau_\text{opt}} \lvert x(t_i + d\tau_\text{opt}) - x(t_{n(i,d)} + d\tau_\text{opt})\rvert
\end{equation}
Since future values are independent of past values for random signals, $E_2(d) \approx 1$ for any   $d$~\cite{cao1997}. In contrast, for deterministic signals, $E_2(d)$ depends on the   embedding dimension $d$, since there must exist some values of $d$ for which $E_2(d) \neq 1$.

\subsection{Route to Intermittency -- Recurrence Plot and Feature Classification}
\label{sec:section_2.3}

Intermittency is a common route to chaos~\cite{ott2002} that manifests as irregularly spaced, alternating intervals of chaotic bursts and steady or periodic behavior. In this state, the dynamical system switches
between two distinct behaviors (also referred to as phases).
Multiple types of intermittency exist, including the three classical types investigated by \citet{pomeau1980}, type X~\cite{price1991}, type V~\cite{he1989, bauer1992}, and a group of chaos–chaos intermittencies~\cite{ott2002}, including on–off~\cite{platt1993} and in–out~\cite{marwan2002, ashwin2001} intermittencies. Each type corresponds to a distinct bifurcation mechanism. For example, type I intermittency occurs near a saddle-node bifurcation, type II near a Hopf bifurcation, and type III near a reverse period-doubling bifurcation~\cite{klimaszewska2009}.

\Citet{marwan2002} demonstrated that recurrence plots (RPs) and recurrence quantification analysis (RQA) can distinguish distinguish between time series with intermittency and other kinds of chaos Mathematically, the recurrence plot is defined by
\begin{equation}
R_{ij} = \Theta (\varepsilon - \lVert x_i - x_j \rVert),
\end{equation}
where $\Theta$ is the Heaviside function, $\varepsilon$ is a predefined threshold, $\lVert \cdot \rVert$ denotes a norm, and $x_i$ represents a point on the attractor in phase space. The recurrence threshold $\varepsilon$ defines the maximum separation between phase space points that qualify as recurrence pairs. They showed that the laminar phases of intermittency correspond to horizontal (and vertical) lines on the RP and that such lines form squares and rectangles~\cite{marwan2005}. The presence of such geometric patterns in the RP indicates the presence of intermittency in the data.

\begin{figure}[htbp]
    \centering
    \includegraphics[width=0.85\textwidth]{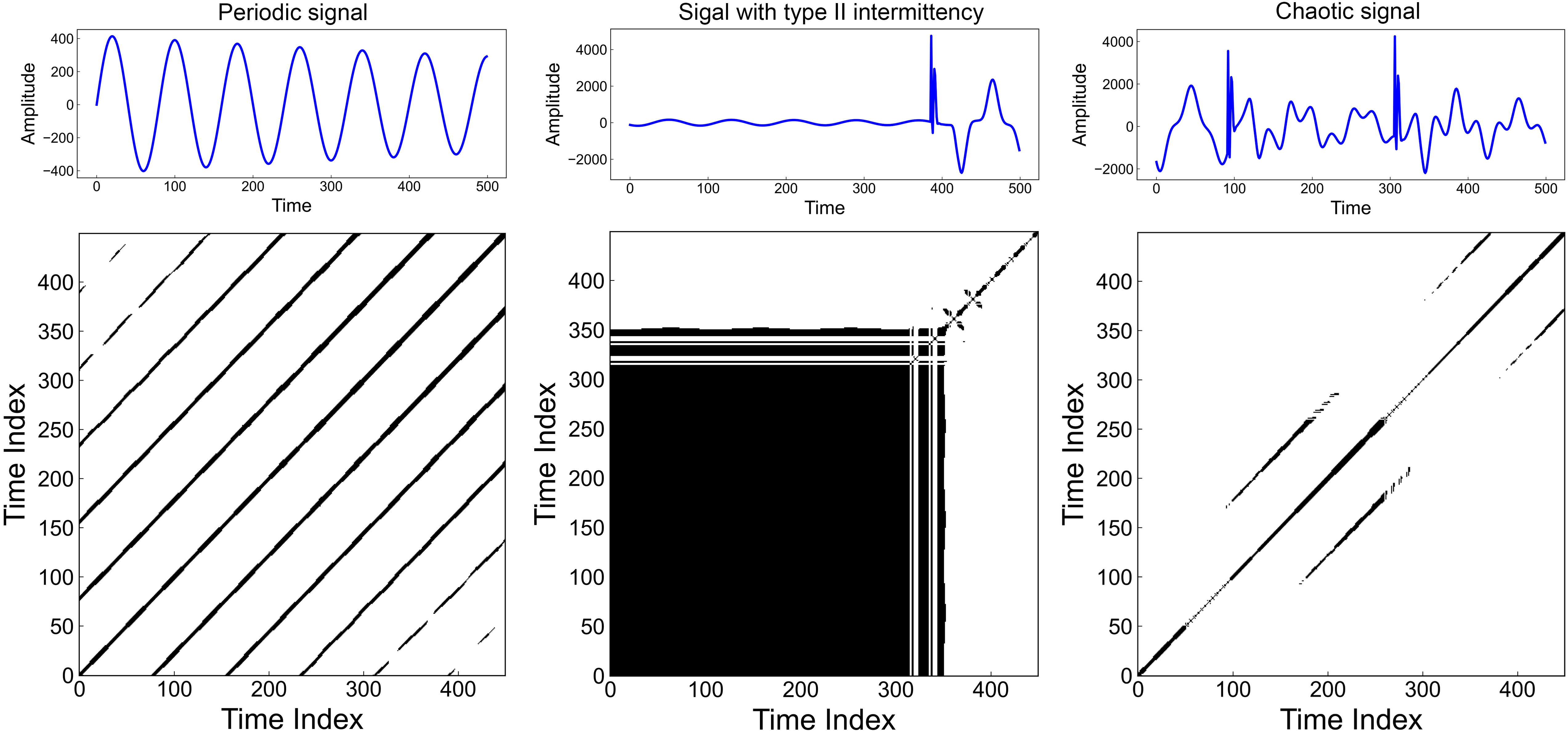}
    \caption{Representative signals and their corresponding recurrence plots showing distinct dynamical regimes. Left: Periodic signal exhibiting regular diagonal patterns in its recurrence plot. Middle: Signal with type II intermittency characterized by laminar phases interrupted by sudden bursts, producing blocked recurrence structures. Right: Chaotic signal displaying irregular fluctuations with scattered recurrence patterns, reflecting the system's underlying unpredictability.}
    \label{fig:figure_3}
\end{figure}

In a RP, black and white points indicate recurrence and non-recurrence, respectively. The points on the main diagonal are black since every point trivially recurs with itself. Several important characteristics of a dynamical system can be inferred from an RP. For example, periodic oscillations, such as limit cycles, produce evenly spaced $45^\circ$ lines parallel to the main diagonal, reflecting the system’s temporal periodicity. In contrast, the RP of a purely random signal exhibits an unstructured distribution of black and white points. The distinguishing features of RPs for periodic, aperiodic, and noisy systems have been analyzed by \citet{marwan2002}. Figure~\ref{fig:figure_3} illustrates these concepts, showing examples of periodic, intermittent, and chaotic signals alongside their corresponding recurrence plots.

\subsubsection*{0--1 Test Classification}

Determining whether a system exhibits regular or chaotic dynamics is a challenge in nonlinear dynamical analysis.  \Citet{gottwald2004} developed a binary diagnostic method, the 0--1 test, specifically designed to distinguish between chaotic and regular dynamics in deterministic systems. This method has demonstrated broad applicability across diverse contexts, including experimental time series, noisy numerical data, quasiperiodically forced systems, strange nonchaotic attractors, Hamiltonian dynamics, non-smooth systems, and fluid mechanics applications. The binary nature of the 0--1 test makes it particularly well-suited for our framework, as it eliminates the ambiguity in threshold-based methods.

\subsubsection*{Intermittency Classification}

Expanding on the approach of \citet{marwan2002}, \citet{klimaszewska2009} observed that different types of intermittency produce distinctive geometric patterns in recurrence plots--including squares, rectangles, and distorted variants--that can be quantitatively characterized. To augment standard RQA, they introduced two additional parameters, $F_a$ and $F_b$, which measure the surface area of RP regions associated with laminar phases.
While this method enables the unambiguous classification of various intermittency types, our primary interest lies in type II intermittency, where the system  hovers near a stable oscillatory state but is occasionally perturbed into erratic behavior due to internal instabilities or stochastic forcing.

The classification framework for the recurrence matrices, outlined in~\cref{alg:classification}, consists of three main steps. Step 1 extracts geometric features from the recurrence matrix, Step 2 applies the 0–1 test for chaos using time-series data, and Step 3 combines these features to assign the appropriate dynamical class.

 \begin{algorithm*}[hbtp!]
\DontPrintSemicolon
\caption{Classification of dynamical regimes using recurrence matrix features and 0-1 test}\label{alg:classification}
\KwIn{Recurrence matrix $M$, time series $X$}
\KwOut{Classification: \emph{Periodic}, \emph{Chaotic}, or \emph{Type II intermittency}}

\BlankLine
\KwSty{Step 1: Recurrence feature extraction}\;
Identify the largest continuous black region in $M$\;
Compute region size $F_a$ (number of black points)\;
Find bounding box of the largest region: $(i_{\text{min}}, j_{\text{min}}), (i_{\text{max}}, j_{\text{max}})$\;
Compute rectangular area: $A_r = (i_{\text{max}} - i_{\text{min}} + 1)\times(j_{\text{max}} - j_{\text{min}} + 1)$\;
Compute area ratio: $F_b = F_a / A_r$\;

\BlankLine
\KwSty{Step 2: 0-1 test for chaos}\;
\If{$X$ is available}{
    Generate random values $c \in [\pi/5, 4\pi/5]$\;
    \ForEach{$c$}{
        Compute translation variables $p(n)$ and $q(n)$:
        \[
        p(n) = \sum_{j=1}^{n} (X_j - \bar{X}) \cos(jc),
        \quad
        q(n) = \sum_{j=1}^{n} (X_j - \bar{X}) \sin(jc)
        \]
        Compute mean square displacement (MSD):
        \[
        D(n) = \frac{1}{N-n} \sum_{j=1}^{N-n} [(p(j+n)-p(j))^2 + (q(j+n)-q(j))^2]
        \]
        Compute $K$-statistic: $K_c = \dfrac{\mathrm{cov}(n, D(n))}{\sqrt{\mathrm{var}(n) \cdot \mathrm{var}(D(n))}}$\;
    }
    $K = \mathrm{median}(K_c)$
}
\BlankLine
\KwSty{Step 3: Classification}\;
\If{$\lvert F_b - 1\rvert < 0.1$}{
    \Return \emph{Type II intermittency}
}
\ElseIf{$K \leq 0.2$}{
    \Return \emph{Periodic}
}
\ElseIf{$K \geq 0.8$}{
    \Return \emph{Chaotic}
}
\Else{
    \Return \emph{Unclassified}
}
\end{algorithm*}

\subsection{Deep Learning Model -- Convolutional Neural Network}
\label{sec:section_2.4}

Our deep learning approach uses Convolutional Neural Networks (CNNs), which are specialized for processing grid-like data such as recurrence matrices. The fundamental operation in a CNN is the convolution, where an input image is filtered by learnable kernels to produce feature maps that highlight relevant patterns in the data. This local filtering mechanism enables CNNs to capture hierarchical features: early layers detect simple elements like edges and textures, while deeper layers combine these to identify complex structures. Spatial dimensionality is typically reduced through pooling operations, which summarize regions of feature maps while preserving key information. CNNs are effective because they learn spatially localized features while significantly reducing the number of parameters compared to fully connected networks, making them well-suited for tasks such as image classification, object detection, and semantic segmentation.
This is particularly well-suited to the recurrence plots used in our analysis, which have  spatial structure that encode visually distinct signatures of underlying dynamical regimes. Details on the CNN architecture are provided in~\cref{supp:cnn}

\section{Benchmark  Systems}

To evaluate the performance and generalizability of our proposed framework, we consider three benchmark dynamical systems: (i) the triad model, (ii) the anisotropic turbulence model, and (iii) the Kolmogorov flow. These models were selected to provide a rigorous test across a hierarchy of complexity, encompassing nonlinear energy transfer, multi-scale intermittency, and the transition to spatio-temporal chaos.

\subsection{Triad Model}

The first test model considered is a canonical three-dimensional system with quadratic, energy-conserving nonlinearity , which we refer to as the \emph{triad model}.  This model captures the three-way nonlinear interactions that drive energy transfer in turbulent flows. Such triadic interactions emerge from the  truncations of  three dominant modes in high-dimensional turbulent system. The model is defined by the following system of equations: 
\begin{align}
d u_1/dt &= -\gamma_1 u_1 + L_{12}u_2 + L_{13}u_3 + Iu_1u_2 + F_1 + \sigma_1\dot{W}_1,\\ d u_2/dt &= -L_{12}u_1 - \gamma_2 u_2 + L_{23}u_3 - Iu_1^2 + \sigma_2\dot{W}_2, \\ 
d u_3 /dt  &= -L_{13}u_1 - L_{23}u_2 - \gamma_3 u_3 + \sigma_3\dot{W}_3.
\end{align}
The stochastic forcing terms  represent unresolved dynamics or external perturbations. Due to the dominant role of nonlinear coupling, the triad model exhibits a wide range of nonlinear and non-Gaussian behaviors, making it an ideal benchmark for evaluating predictive modeling strategies.   Despite its nonlinear structure, the triad model equilibrium statistics are analytically tractable under special  parameter configurations. Further details are provided in~\cref{supp:doc2}.

\subsection{Anisotropic Turbulent Flow Model}

The second test model is a conceptual, low-dimensional stochastic system designed to capture key statistical properties of anisotropic turbulence through energy-conserving wave–mean-flow interactions and stochastic forcing on the fluctuations~\cite{majda2014}. The model consists of a mean scalar variable $\bar u$, representing the largest  scales (mean flow), and a set of small-scale fluctuations $\mathbf{u}'=(u_1',u_2',\ldots,u_K')$, with the total turbulent field given by
$u(t)=\bar{u}(t)+\sum_{k}u_k'(t)$.
The system evolves according to 
\begin{align}
d\bar{u}/dt &=-\bar{d}\bar{u}+\gamma\sum_{k=1}^{K}(u_k')^2-\bar{\alpha}\bar{u}^3+\bar{F}\\ 
d u_k'/dt &=-d_ku_k'-\gamma\bar{u}u_k'+\sigma_k\dot{W}_k,\quad\text{for } 1 \leq k \leq K,
\end{align}
where the nonlinear terms conserve the total energy $E(\bar{u},\mathbf{u}')= \tfrac{1}{2}\bar{u}^2+ \tfrac{1}{2}\sum_{k}u_k'^2$. 
A key feature of this model is that large scales can destabilize smaller scales when $-d_k-\gamma\bar{u}>0$, that is, $\bar{u}<-\frac{d_k}{\gamma}$, triggering intermittent instabilities that increase small-scale energy and feed back to large scales. Despite its simplicity, the model captures essential statistical features of anisotropic turbulence, including chaotic mean-flow behavior with a sub-Gaussian probability distribution, along with decreasing energy and correlation times at smaller scales. Large-scale fluctuations exhibit nearly Gaussian PDFs, whereas smaller-scale fluctuations display fat-tailed, non-Gaussian PDFs--a hallmark of intermittency, where modes with small variance exhibit relatively frequent extreme events that impact the mean flow. Further details are provided in~\cref{supp:doc3}.

\subsection{Kolmogorov Flow Model}

The Kolmogorov flow is a classic model in  fluid dynamics,  characterized by  sinusoidal forcing,
$F(y) = F_0 \sin(ny)$, which generates a unidirectional, laminar base flow,
$U(y) = U_0 \sin(ny)$, with a spatially periodic structure. Originally proposed  as a theoretical framework for investigating the transition to turbulence, this system is governed by the Navier–Stokes equations:
\begin{equation}
\partial_t \mathbf{u} + (\mathbf{u} \cdot \nabla)\mathbf{u} = -\nabla p + \nu \nabla^2 \mathbf{u} + \mathbf{F}
\end{equation}
with the forcing term   $\mathbf{F} = (F_0 \sin(ny), 0)$.
At a critical Reynolds number, $Re = U_0 / (\nu n)$, the flow undergoes a symmetry-breaking instability that gives rise to coherent vortical structures with characteristic vorticity $\omega = \nabla \times \mathbf{u}$, eventually transitioning into chaotic regimes. Numerical and experimental studies have established Kolmogorov flow as an ideal test bed for studying fundamental aspects of hydrodynamic stability, pattern formation, and the statistical properties of two-dimensional turbulence. More details are provided in~\cref{supp:doc4}.

\section{Results and Discussion}

\subsection{CNN Prediction Performance}

\begin{figure}[htbp]
    \centering
    \includegraphics[width=0.85\linewidth]{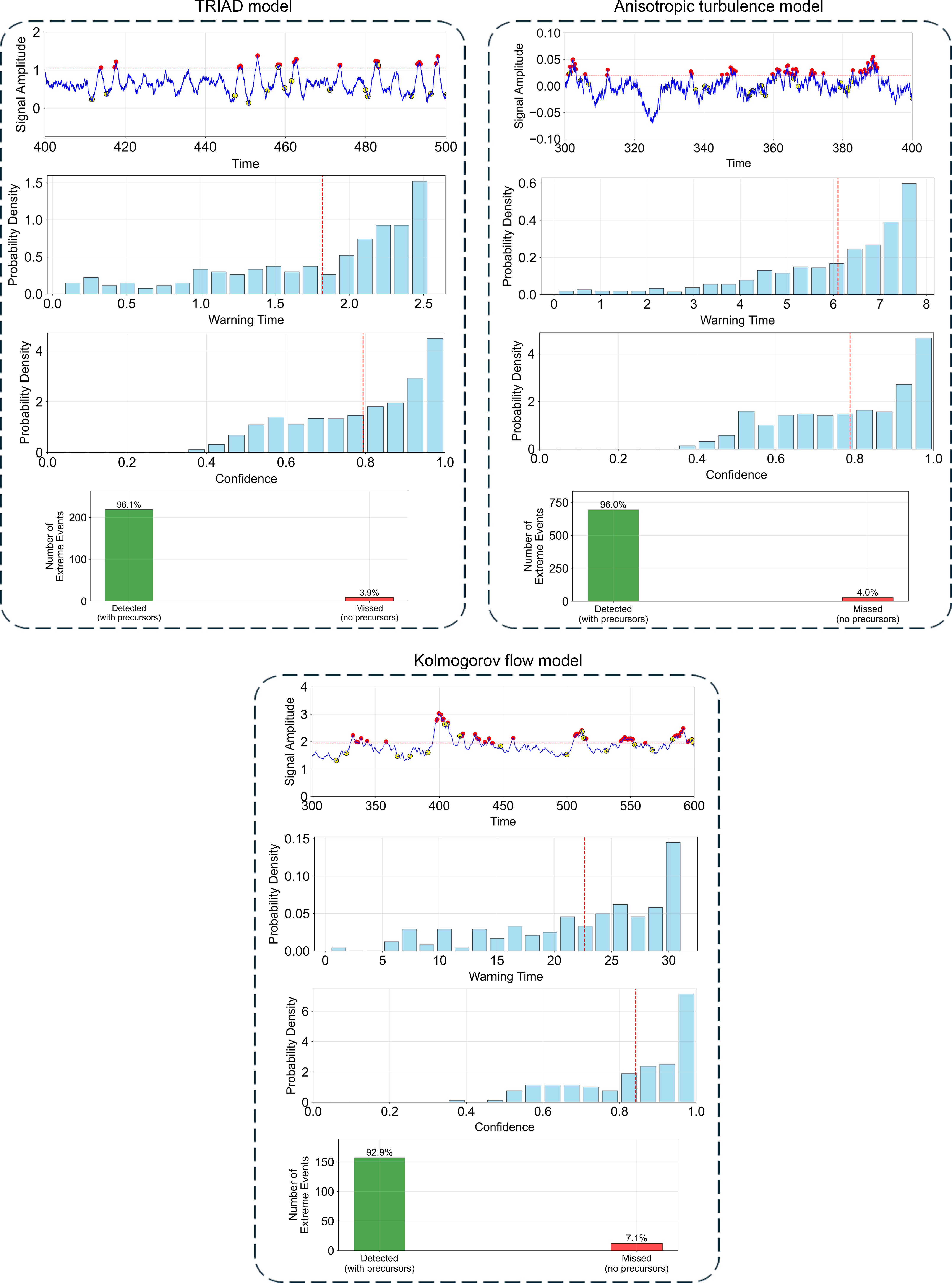}
    \caption{Performance evaluation of the precursor detection framework across three distinct dynamical systems: TRIAD model (left), anisotropic turbulence model (right), and Kolmogorov flow model (bottom). For each model, the figure displays the time series with extreme events (red dots) and their identified precursors (yellow dots), warning time distribution, prediction confidence distribution, and detection performance metrics.}
    \label{fig:figure_4}
\end{figure}

To evaluate the predictive performance of the CNN-based precursor detection framework, we apply the method to three canonical systems: (1) the triad model, (2) the anisotropic turbulence model, and (3) the Kolmogorov flow model. Figure~\ref{fig:figure_4} summarizes the results.

Extreme events are identified by  local maxima detection using the signal amplitude. An event is classified as `extreme' when 
\begin{equation}
 u(\hat{t}_i) > \mu + n \sigma   
\end{equation}
where $n$ is a constant, $\mu = \mathbb{E}[u(t)]$ is the mean, and $\sigma^2 =  \mathbb{V}[u(t)]$ is the variance.
For each extreme event at time $t_{\text{e}}$, a precursor is defined as  the earliest type II  pattern identified  by the CNN at time $t_{\text{p}}$ such that $t_{\text{p}} < t_{\text{e}}$ and $t_{\text{e}} - t_{\text{p}} \leq \tau_{\text{max}}$, where $\tau_{\text{max}}$ denotes  the maximum lookback window.

From these precursor-extreme event pairs, we compute the warning time
\begin{equation}
\Delta t_i = t_{\text{e},i} - t_{\text{p},i}   
\end{equation}
for each pair. The probability density $p(\Delta t)$ is then estimated using a normalized histogram. The statistical  moments of this distribution characterize the system's predictability.

To assess the reliability of CNN-generated type II predictions, we analyze the distribution of their associated confidence scores $c \in [0,1]$. The probability density $p(c)$ is estimated using bin intervals $\Delta c_k = [c_k, c_{k+1}]$, with bin-specific success rates defined $R_k = N_{\text{s},k}/N_{\text{total},k}$, where $N_{\text{s},k}$ is the number of successful predictions in bin $k$ and $N_{\text{total},k}$ is the total of predictions in bin $k$.
 We further compute statistical measures for the subsets of successful and unsuccessful predictions separately: $\mu_{\text{s}} = \mathbb{E}[c \mid c \in \mathcal{C}_{\text{s}}]$ and $\mu_{\text{f}} = \mathbb{E}[c \mid c \in \mathcal{C}_{\text{f}}]$, where $\mathcal{C}_{\text{s}}$ and $\mathcal{C}_{\text{f}}$  represent the sets of confidence scores corresponding to successful and unsuccessful predictions, respectively.

Finally, we quantify the method's detection efficiency by comparing the number of extreme events with identified precursors, $N_{\text{d}}$, against those without precursors,  $N_{\text{m}}$. The detection rate is defined as $\eta = {N_{\text{d}}}/({N_{\text{d}} + N_{\text{m}}})$, representing the fraction of extreme events successfully predicted by the model. An extreme event is considered detected if at least one type II prediction occurs within the lookback window, $\tau_{\text{max}}$, preceding the event.

\subsection{Benchmark System Performance}

\paragraph{Triad Model}  
The triad model showed strong predictive performance. Our method correctly identified 96.1\% of extreme events (219 out of 228), demonstrating the efficacy of the recurrence matrix–CNN approach in detecting precursors that precede extreme excursions. Temporal statistics indicated a mean lead time of 1.82 time units (median: 2.06), with a standard deviation of 0.66 and a range spanning from 0.09 to 2.55 time units. The distribution of prediction confidence was skewed towards higher values, with $55.8\%$ of the predictions falling within the $[0.8, 1.0]$ interval and $37.0\%$ in the upper decile $[0.9, 1.0]$.

\paragraph{Anisotropic Turbulence Model}  
The anisotropic turbulent flow model showed similarly   high predictive performance, with our framework correctly identifying precursors for $96.0\%$ of extreme events (695 out of 724 occurrences). This system demonstrated intermediate   lead times, with a mean of 6.10 time units (median: 6.66) and moderate variability ($\sigma = 1.70$), ranging from 0.11 to 7.86 time units. Confidence analysis indicated that 52.9\% of predictions fell within the high-confidence interval $[0.8, 1.0]$, with 36.9\% in the top decile  $[0.9, 1.0]$, comparable to the triad model results despite the increased complexity of the dynamical system.

\paragraph{Kolmogorov Flow}
The Kolmogorov flow provides strong evidence of our framework's generalizability, with 92.9\% of extreme events correctly anticipated (157 out of 169 occurrences). This canonical flow exhibited the longest prediction horizon among the three systems, with a mean lead time of 22.66 time units (median: 24.75) and greater variability ($\sigma = 7.56 $), ranging from 0.57 to 31.31 time units. Note that time units are system-specific, reflecting intrinsic dynamical timescales rather than allowing direct cross-system comparisons. In particular, this system showed the highest confidence levels, with 69.3\% of the predictions falling within the high confidence interval $[0.8, 1.0]$, and nearly half (48.1\%) in the upper decile $[0.9, 1.0]$.

\section{Conclusions}

In this study, we have developed a framework that combines phase-space reconstruction, recurrence matrices, and convolutional neural networks to predict extreme events in turbulent dynamical systems. Testing in three distinct systems, the triad model, the anisotropic turbulent flow, and the Kolmogorov flow, yielded detection rates of 96.1\%, 96.0\%, and 92.9\%, respectively.

Three features distinguish this approach from existing methods. The threshold-free classification eliminates the parameter selection bias inherent in traditional techniques. Training requires only \textasciitilde750 recurrence matrices, substantially fewer than conventional machine learning approaches for chaotic systems. Most significantly, the method generalizes across systems with fundamentally different physics, indicating that recurrence matrices encode universal precursor patterns independent of specific system dynamics.

These results suggest that the integration of dynamical system theory with pattern recognition can reliably detect extreme event precursors in turbulent flows. The method's generalizability and computational efficiency make it particularly suitable for real-time applications in engineering systems where early warning of extreme events is critical. Future work explore its extension to multivariate time-series data,  analyze sensitivity to noise and parameter uncertainty, and applications to higher-dimensional and experimental systems.

\section*{Acknowledgments}
The authors acknowledge financial support from Alberta Innovates and the Natural Sciences and Engineering Research Council of Canada (NSERC).

\bibliographystyle{elsarticle-num-names}
\bibliography{refs}

\begin{thebibliography}{39}
\expandafter\ifx\csname natexlab\endcsname\relax\def\natexlab#1{#1}\fi
\providecommand{\url}[1]{\texttt{#1}}
\providecommand{\href}[2]{#2}
\providecommand{\path}[1]{#1}
\providecommand{\DOIprefix}{doi:}
\providecommand{\ArXivprefix}{arXiv:}
\providecommand{\URLprefix}{URL: }
\providecommand{\Pubmedprefix}{pmid:}
\providecommand{\doi}[1]{\href{http://dx.doi.org/#1}{\path{#1}}}
\providecommand{\Pubmed}[1]{\href{pmid:#1}{\path{#1}}}
\providecommand{\bibinfo}[2]{#2}
\ifx\xfnm\relax \def\xfnm[#1]{\unskip,\space#1}\fi
\bibitem[{Davini et~al.(2012)Davini, Cagnazzo, Gualdi, and Navarra}]{davini2012}
\bibinfo{author}{P.~Davini}, \bibinfo{author}{C.~Cagnazzo}, \bibinfo{author}{S.~Gualdi}, \bibinfo{author}{A.~Navarra},
\newblock \bibinfo{title}{Bidimensional {{Diagnostics}}, {{Variability}}, and {{Trends}} of {{Northern Hemisphere Blocking}}},
\newblock \bibinfo{journal}{J. Clim.} \bibinfo{volume}{25} (\bibinfo{year}{2012}) \bibinfo{pages}{6496--6509}. \DOIprefix\doi{10.1175/JCLI-D-12-00032.1}.
\bibitem[{Jim{\'e}nez and Simens(2001)}]{jimenez2001}
\bibinfo{author}{J.~Jim{\'e}nez}, \bibinfo{author}{M.~P. Simens},
\newblock \bibinfo{title}{Low-dimensional dynamics of a turbulent wall flow},
\newblock \bibinfo{journal}{J. Fluid Mech.} \bibinfo{volume}{435} (\bibinfo{year}{2001}) \bibinfo{pages}{81--91}. \DOIprefix\doi{10.1017/S0022112001004050}.
\bibitem[{Ishihara et~al.(2007)Ishihara, Kaneda, Yokokawa, Itakura, and Uno}]{ishihara2007}
\bibinfo{author}{T.~Ishihara}, \bibinfo{author}{Y.~Kaneda}, \bibinfo{author}{M.~Yokokawa}, \bibinfo{author}{K.~Itakura}, \bibinfo{author}{A.~Uno},
\newblock \bibinfo{title}{Small-scale statistics in high-resolution direct numerical simulation of turbulence: {{Reynolds}} number dependence of one-point velocity gradient statistics},
\newblock \bibinfo{journal}{J. Fluid Mech.} \bibinfo{volume}{592} (\bibinfo{year}{2007}) \bibinfo{pages}{335--366}. \DOIprefix\doi{10.1017/S0022112007008531}.
\bibitem[{Iyer and Ceccio(2002)}]{iyer2002}
\bibinfo{author}{C.~O. Iyer}, \bibinfo{author}{S.~L. Ceccio},
\newblock \bibinfo{title}{The influence of developed cavitation on the flow of a turbulent shear layer},
\newblock \bibinfo{journal}{Physics of Fluids} \bibinfo{volume}{14} (\bibinfo{year}{2002}) \bibinfo{pages}{3414--3431}. \DOIprefix\doi{10.1063/1.1501541}.
\bibitem[{Schmidt and Schmid(2019)}]{schmidt2019}
\bibinfo{author}{O.~T. Schmidt}, \bibinfo{author}{P.~J. Schmid},
\newblock \bibinfo{title}{A conditional space--time {{POD}} formalism for intermittent and rare events: Example of acoustic bursts in turbulent jets},
\newblock \bibinfo{journal}{J. Fluid Mech.} \bibinfo{volume}{867} (\bibinfo{year}{2019}) \bibinfo{pages}{R2}. \DOIprefix\doi{10.1017/jfm.2019.200}.
\bibitem[{{Modarres-Sadeghi} et~al.(2010){Modarres-Sadeghi}, Mukundan, Dahl, Hover, and Triantafyllou}]{modarres-sadeghi2010}
\bibinfo{author}{Y.~{Modarres-Sadeghi}}, \bibinfo{author}{H.~Mukundan}, \bibinfo{author}{J.~M. Dahl}, \bibinfo{author}{F.~S. Hover}, \bibinfo{author}{M.~S. Triantafyllou},
\newblock \bibinfo{title}{The effect of higher harmonic forces on fatigue life of marine risers},
\newblock \bibinfo{journal}{Journal of Sound and Vibration} \bibinfo{volume}{329} (\bibinfo{year}{2010}) \bibinfo{pages}{43--55}. \DOIprefix\doi{10.1016/j.jsv.2009.07.024}.
\bibitem[{Belenky et~al.(2019)Belenky, Glotzer, Pipiras, and Sapsis}]{belenky2019}
\bibinfo{author}{V.~Belenky}, \bibinfo{author}{D.~Glotzer}, \bibinfo{author}{V.~Pipiras}, \bibinfo{author}{T.~P. Sapsis},
\newblock \bibinfo{title}{Distribution tail structure and extreme value analysis of constrained piecewise linear oscillators},
\newblock \bibinfo{journal}{Probabilistic Engineering Mechanics} \bibinfo{volume}{57} (\bibinfo{year}{2019}) \bibinfo{pages}{1--13}. \DOIprefix\doi{10.1016/j.probengmech.2019.04.001}.
\bibitem[{Mohamad and Sapsis(2016)}]{mohamad2016}
\bibinfo{author}{M.~A. Mohamad}, \bibinfo{author}{T.~P. Sapsis},
\newblock \bibinfo{title}{Probabilistic response and rare events in {{Mathieu}}'s equation under correlated parametric excitation},
\newblock \bibinfo{journal}{Ocean Engineering} \bibinfo{volume}{120} (\bibinfo{year}{2016}) \bibinfo{pages}{289--297}. \DOIprefix\doi{10.1016/j.oceaneng.2016.03.008}.
\bibitem[{Sapsis and Majda(2013)}]{sapsis2013}
\bibinfo{author}{T.~P. Sapsis}, \bibinfo{author}{A.~J. Majda},
\newblock \bibinfo{title}{A statistically accurate modified quasilinear {{Gaussian}} closure for uncertainty quantification in turbulent dynamical systems},
\newblock \bibinfo{journal}{Physica D: Nonlinear Phenomena} \bibinfo{volume}{252} (\bibinfo{year}{2013}) \bibinfo{pages}{34--45}. \DOIprefix\doi{10.1016/j.physd.2013.02.009}.
\bibitem[{Mohamad et~al.(2016)Mohamad, Cousins, and Sapsis}]{mohamad2016a}
\bibinfo{author}{M.~A. Mohamad}, \bibinfo{author}{W.~Cousins}, \bibinfo{author}{T.~P. Sapsis},
\newblock \bibinfo{title}{A probabilistic decomposition-synthesis method for the quantification of rare events due to internal instabilities},
\newblock \bibinfo{journal}{Journal of Computational Physics} \bibinfo{volume}{322} (\bibinfo{year}{2016}) \bibinfo{pages}{288--308}. \DOIprefix\doi{10.1016/j.jcp.2016.06.047}.
\bibitem[{Mohamad and Sapsis(2018)}]{mohamad2018}
\bibinfo{author}{M.~A. Mohamad}, \bibinfo{author}{T.~P. Sapsis},
\newblock \bibinfo{title}{Sequential sampling strategy for extreme event statistics in nonlinear dynamical systems},
\newblock \bibinfo{journal}{Proc. Natl. Acad. Sci. U.S.A.} \bibinfo{volume}{115} (\bibinfo{year}{2018}) \bibinfo{pages}{11138--11143}. \DOIprefix\doi{10.1073/pnas.1813263115}.
\bibitem[{Doan et~al.(2021)Doan, Polifke, and Magri}]{doan2021}
\bibinfo{author}{N.~a.~K. Doan}, \bibinfo{author}{W.~Polifke}, \bibinfo{author}{L.~Magri},
\newblock \bibinfo{title}{Short- and long-term predictions of chaotic flows and extreme events: A physics-constrained reservoir computing approach},
\newblock \bibinfo{journal}{Proc. R. Soc. Math. Phys. Eng. Sci.} \bibinfo{volume}{477} (\bibinfo{year}{2021}) \bibinfo{pages}{20210135}. \DOIprefix\doi{10.1098/rspa.2021.0135}.
\bibitem[{Blonigan et~al.(2019)Blonigan, Farazmand, and Sapsis}]{blonigan2019}
\bibinfo{author}{P.~J. Blonigan}, \bibinfo{author}{M.~Farazmand}, \bibinfo{author}{T.~P. Sapsis},
\newblock \bibinfo{title}{Are extreme dissipation events predictable in turbulent fluid flows?},
\newblock \bibinfo{journal}{Phys. Rev. Fluids} \bibinfo{volume}{4} (\bibinfo{year}{2019}) \bibinfo{pages}{044606}. \DOIprefix\doi{10.1103/PhysRevFluids.4.044606}.
\bibitem[{Dematteis et~al.(2018)Dematteis, Grafke, and {Vanden-Eijnden}}]{dematteis2018}
\bibinfo{author}{G.~Dematteis}, \bibinfo{author}{T.~Grafke}, \bibinfo{author}{E.~{Vanden-Eijnden}},
\newblock \bibinfo{title}{Rogue waves and large deviations in deep sea},
\newblock \bibinfo{journal}{Proc. Natl. Acad. Sci.} \bibinfo{volume}{115} (\bibinfo{year}{2018}) \bibinfo{pages}{855--860}. \DOIprefix\doi{10.1073/pnas.1710670115}.
\bibitem[{Vlachas et~al.(2018)Vlachas, Byeon, Wan, Sapsis, and Koumoutsakos}]{vlachas2018}
\bibinfo{author}{P.~R. Vlachas}, \bibinfo{author}{W.~Byeon}, \bibinfo{author}{Z.~Y. Wan}, \bibinfo{author}{T.~P. Sapsis}, \bibinfo{author}{P.~Koumoutsakos},
\newblock \bibinfo{title}{Data-driven forecasting of high-dimensional chaotic systems with long short-term memory networks},
\newblock \bibinfo{journal}{Proc. R. Soc. Math. Phys. Eng. Sci.} \bibinfo{volume}{474} (\bibinfo{year}{2018}) \bibinfo{pages}{20170844}. \DOIprefix\doi{10.1098/rspa.2017.0844}.
\bibitem[{Srinivasan et~al.(2019)Srinivasan, Guastoni, Azizpour, Schlatter, and Vinuesa}]{Srinivasan2019}
\bibinfo{author}{P.~A. Srinivasan}, \bibinfo{author}{L.~Guastoni}, \bibinfo{author}{H.~Azizpour}, \bibinfo{author}{P.~Schlatter}, \bibinfo{author}{R.~Vinuesa},
\newblock \bibinfo{title}{Predictions of turbulent shear flows using deep neural networks},
\newblock \bibinfo{journal}{Phys. Rev. Fluids} \bibinfo{volume}{4} (\bibinfo{year}{2019}) \bibinfo{pages}{054603}. \DOIprefix\doi{10.1103/PhysRevFluids.4.054603}.
\bibitem[{Pathak et~al.(2018{\natexlab{a}})Pathak, Wikner, Fussell, Chandra, Hunt, Girvan, and Ott}]{pathak2018}
\bibinfo{author}{J.~Pathak}, \bibinfo{author}{A.~Wikner}, \bibinfo{author}{R.~Fussell}, \bibinfo{author}{S.~Chandra}, \bibinfo{author}{B.~R. Hunt}, \bibinfo{author}{M.~Girvan}, \bibinfo{author}{E.~Ott},
\newblock \bibinfo{title}{Hybrid forecasting of chaotic processes: {{Using}} machine learning in conjunction with a knowledge-based model},
\newblock \bibinfo{journal}{Chaos} \bibinfo{volume}{28} (\bibinfo{year}{2018}{\natexlab{a}}) \bibinfo{pages}{041101}. \DOIprefix\doi{10.1063/1.5028373}.
\bibitem[{Pathak et~al.(2018{\natexlab{b}})Pathak, Hunt, Girvan, Lu, and Ott}]{pathak2018a}
\bibinfo{author}{J.~Pathak}, \bibinfo{author}{B.~Hunt}, \bibinfo{author}{M.~Girvan}, \bibinfo{author}{Z.~Lu}, \bibinfo{author}{E.~Ott},
\newblock \bibinfo{title}{Model-{{Free Prediction}} of {{Large Spatiotemporally Chaotic Systems}} from {{Data}}: {{A Reservoir Computing Approach}}},
\newblock \bibinfo{journal}{Phys. Rev. Lett.} \bibinfo{volume}{120} (\bibinfo{year}{2018}{\natexlab{b}}) \bibinfo{pages}{024102}. \DOIprefix\doi{10.1103/PhysRevLett.120.024102}.
\bibitem[{Doan et~al.(2019)Doan, Polifke, and Magri}]{doan2019}
\bibinfo{author}{N.~A.~K. Doan}, \bibinfo{author}{W.~Polifke}, \bibinfo{author}{L.~Magri},
\newblock \bibinfo{title}{Physics-{{Informed Echo State Networks}} for {{Chaotic Systems Forecasting}}},
\newblock in: \bibinfo{editor}{J.~M.~F. Rodrigues}, \bibinfo{editor}{P.~J.~S. Cardoso}, \bibinfo{editor}{J.~Monteiro}, \bibinfo{editor}{R.~Lam}, \bibinfo{editor}{V.~V. Krzhizhanovskaya}, \bibinfo{editor}{M.~H. Lees}, \bibinfo{editor}{J.~J. Dongarra}, \bibinfo{editor}{P.~M. Sloot} (Eds.), \bibinfo{booktitle}{Comput. {{Sci}}. -- {{ICCS}} 2019}, \bibinfo{publisher}{Springer International Publishing}, \bibinfo{address}{Cham}, \bibinfo{year}{2019}, pp. \bibinfo{pages}{192--198}. \DOIprefix\doi{10.1007/978-3-030-22747-0_15}.
\bibitem[{Huhn and Magri(2020)}]{huhn2020}
\bibinfo{author}{F.~Huhn}, \bibinfo{author}{L.~Magri},
\newblock \bibinfo{title}{Learning {{Ergodic Averages}} in {{Chaotic Systems}}},
\newblock in: \bibinfo{editor}{V.~V. Krzhizhanovskaya}, \bibinfo{editor}{G.~Z{\'a}vodszky}, \bibinfo{editor}{M.~H. Lees}, \bibinfo{editor}{J.~J. Dongarra}, \bibinfo{editor}{P.~M.~A. Sloot}, \bibinfo{editor}{S.~Brissos}, \bibinfo{editor}{J.~Teixeira} (Eds.), \bibinfo{booktitle}{Comput. {{Sci}}. -- {{ICCS}} 2020}, \bibinfo{publisher}{Springer International Publishing}, \bibinfo{address}{Cham}, \bibinfo{year}{2020}, pp. \bibinfo{pages}{124--132}. \DOIprefix\doi{10.1007/978-3-030-50433-5_10}.
\bibitem[{Wan et~al.(2018)Wan, Vlachas, Koumoutsakos, and Sapsis}]{wan2018}
\bibinfo{author}{Z.~Y. Wan}, \bibinfo{author}{P.~Vlachas}, \bibinfo{author}{P.~Koumoutsakos}, \bibinfo{author}{T.~Sapsis},
\newblock \bibinfo{title}{Data-assisted reduced-order modeling of extreme events in complex dynamical systems},
\newblock \bibinfo{journal}{PLOS ONE} \bibinfo{volume}{13} (\bibinfo{year}{2018}) \bibinfo{pages}{e0197704}. \DOIprefix\doi{10.1371/journal.pone.0197704}.
\bibitem[{Farazmand and Sapsis(2017)}]{farazmand2017}
\bibinfo{author}{M.~Farazmand}, \bibinfo{author}{T.~P. Sapsis},
\newblock \bibinfo{title}{A variational approach to probing extreme events in turbulent dynamical systems},
\newblock \bibinfo{journal}{Sci. Adv.} \bibinfo{volume}{3} (\bibinfo{year}{2017}) \bibinfo{pages}{e1701533}. \DOIprefix\doi{10.1126/sciadv.1701533}.
\bibitem[{Soons et~al.(2025)Soons, Grafke, and Dijkstra}]{soons2025}
\bibinfo{author}{J.~Soons}, \bibinfo{author}{T.~Grafke}, \bibinfo{author}{H.~A. Dijkstra},
\newblock \bibinfo{title}{Most likely noise-induced tipping of the overturning circulation in a two-dimensional {{Boussinesq}} fluid model},
\newblock \bibinfo{journal}{J. Fluid Mech.} \bibinfo{volume}{1009} (\bibinfo{year}{2025}) \bibinfo{pages}{A53}. \DOIprefix\doi{10.1017/jfm.2025.248}.
\bibitem[{Matveev and Culick(2003)}]{matveev2003}
\bibinfo{author}{K.~I. Matveev}, \bibinfo{author}{F.~E.~C. Culick},
\newblock \bibinfo{title}{A model for combustion instability involving vortex shedding},
\newblock \bibinfo{journal}{Combust. Sci. Technol.} \bibinfo{volume}{175} (\bibinfo{year}{2003}) \bibinfo{pages}{1059--1083}. \DOIprefix\doi{10.1080/00102200302349}.
\bibitem[{Takens(1981)}]{takens1981}
\bibinfo{author}{F.~Takens},
\newblock \bibinfo{title}{Detecting strange attractors in turbulence},
\newblock in: \bibinfo{editor}{D.~Rand}, \bibinfo{editor}{L.-S. Young} (Eds.), \bibinfo{booktitle}{Dyn. {{Syst}}. {{Turbul}}. {{Warwick}} 1980}, \bibinfo{publisher}{Springer}, \bibinfo{address}{Berlin, Heidelberg}, \bibinfo{year}{1981}, pp. \bibinfo{pages}{366--381}. \DOIprefix\doi{10.1007/BFb0091924}.
\bibitem[{Abarbanel et~al.(1993)Abarbanel, Brown, Sidorowich, and Tsimring}]{abarbanel1993}
\bibinfo{author}{H.~D.~I. Abarbanel}, \bibinfo{author}{R.~Brown}, \bibinfo{author}{J.~J. Sidorowich}, \bibinfo{author}{L.~S. Tsimring},
\newblock \bibinfo{title}{The analysis of observed chaotic data in physical systems},
\newblock \bibinfo{journal}{Rev. Mod. Phys.} \bibinfo{volume}{65} (\bibinfo{year}{1993}) \bibinfo{pages}{1331--1392}. \DOIprefix\doi{10.1103/RevModPhys.65.1331}.
\bibitem[{Cao(1997)}]{cao1997}
\bibinfo{author}{L.~Cao},
\newblock \bibinfo{title}{Practical method for determining the minimum embedding dimension of a scalar time series},
\newblock \bibinfo{journal}{Physica D: Nonlinear Phenomena} \bibinfo{volume}{110} (\bibinfo{year}{1997}) \bibinfo{pages}{43--50}. \DOIprefix\doi{10.1016/S0167-2789(97)00118-8}.
\bibitem[{Ott(2002)}]{ott2002}
\bibinfo{author}{E.~Ott}, \bibinfo{title}{Chaos in {{Dynamical Systems}}}, \bibinfo{edition}{2} ed., \bibinfo{publisher}{Cambridge University Press}, \bibinfo{address}{Cambridge}, \bibinfo{year}{2002}. \DOIprefix\doi{10.1017/CBO9780511803260}.
\bibitem[{Pomeau and Manneville(1980)}]{pomeau1980}
\bibinfo{author}{Y.~Pomeau}, \bibinfo{author}{P.~Manneville},
\newblock \bibinfo{title}{Intermittent transition to turbulence in dissipative dynamical systems},
\newblock \bibinfo{journal}{Commun.Math. Phys.} \bibinfo{volume}{74} (\bibinfo{year}{1980}) \bibinfo{pages}{189--197}. \DOIprefix\doi{10.1007/BF01197757}.
\bibitem[{Price and Mullin(1991)}]{price1991}
\bibinfo{author}{T.~J. Price}, \bibinfo{author}{T.~Mullin},
\newblock \bibinfo{title}{An experimental observation of a new type of intermittency},
\newblock \bibinfo{journal}{Physica D: Nonlinear Phenomena} \bibinfo{volume}{48} (\bibinfo{year}{1991}) \bibinfo{pages}{29--52}. \DOIprefix\doi{10.1016/0167-2789(91)90050-J}.
\bibitem[{He et~al.(1989)He, Wang, Shi, Yang, Chao, and Zhang}]{he1989}
\bibinfo{author}{D.-R. He}, \bibinfo{author}{D.-k. Wang}, \bibinfo{author}{K.-J. Shi}, \bibinfo{author}{C.-h. Yang}, \bibinfo{author}{L.-y. Chao}, \bibinfo{author}{J.-y. Zhang},
\newblock \bibinfo{title}{Critical behavior of dynamical systems described by the inverse circle map},
\newblock \bibinfo{journal}{Physics Letters A} \bibinfo{volume}{136} (\bibinfo{year}{1989}) \bibinfo{pages}{363--368}. \DOIprefix\doi{10.1016/0375-9601(89)90416-7}.
\bibitem[{Bauer et~al.(1992)Bauer, Habip, He, and Martienssen}]{bauer1992}
\bibinfo{author}{M.~Bauer}, \bibinfo{author}{S.~Habip}, \bibinfo{author}{D.~R. He}, \bibinfo{author}{W.~Martienssen},
\newblock \bibinfo{title}{New type of intermittency in discontinuous maps},
\newblock \bibinfo{journal}{Phys. Rev. Lett.} \bibinfo{volume}{68} (\bibinfo{year}{1992}) \bibinfo{pages}{1625--1628}. \DOIprefix\doi{10.1103/PhysRevLett.68.1625}.
\bibitem[{Platt et~al.(1993)Platt, Spiegel, and Tresser}]{platt1993}
\bibinfo{author}{N.~Platt}, \bibinfo{author}{E.~A. Spiegel}, \bibinfo{author}{C.~Tresser},
\newblock \bibinfo{title}{On-off intermittency: {{A}} mechanism for bursting},
\newblock \bibinfo{journal}{Phys. Rev. Lett.} \bibinfo{volume}{70} (\bibinfo{year}{1993}) \bibinfo{pages}{279--282}. \DOIprefix\doi{10.1103/PhysRevLett.70.279}.
\bibitem[{Marwan et~al.(2002)Marwan, Wessel, Meyerfeldt, Schirdewan, and Kurths}]{marwan2002}
\bibinfo{author}{N.~Marwan}, \bibinfo{author}{N.~Wessel}, \bibinfo{author}{U.~Meyerfeldt}, \bibinfo{author}{A.~Schirdewan}, \bibinfo{author}{J.~Kurths},
\newblock \bibinfo{title}{Recurrence-plot-based measures of complexity and their application to heart-rate-variability data},
\newblock \bibinfo{journal}{Phys. Rev. E} \bibinfo{volume}{66} (\bibinfo{year}{2002}) \bibinfo{pages}{026702}. \DOIprefix\doi{10.1103/PhysRevE.66.026702}.
\bibitem[{Ashwin et~al.(2001)Ashwin, Covas, and Tavakol}]{ashwin2001}
\bibinfo{author}{P.~Ashwin}, \bibinfo{author}{E.~Covas}, \bibinfo{author}{R.~Tavakol},
\newblock \bibinfo{title}{Influence of noise on scalings for in-out intermittency},
\newblock \bibinfo{journal}{Phys. Rev. E} \bibinfo{volume}{64} (\bibinfo{year}{2001}) \bibinfo{pages}{066204}. \DOIprefix\doi{10.1103/PhysRevE.64.066204}.
\bibitem[{Klimaszewska and {\.Z}ebrowski(2009)}]{klimaszewska2009}
\bibinfo{author}{K.~Klimaszewska}, \bibinfo{author}{J.~J. {\.Z}ebrowski},
\newblock \bibinfo{title}{Detection of the type of intermittency using characteristic patterns in recurrence plots},
\newblock \bibinfo{journal}{Phys. Rev. E} \bibinfo{volume}{80} (\bibinfo{year}{2009}) \bibinfo{pages}{026214}. \DOIprefix\doi{10.1103/PhysRevE.80.026214}.
\bibitem[{Marwan and Kurths(2005)}]{marwan2005}
\bibinfo{author}{N.~Marwan}, \bibinfo{author}{J.~Kurths},
\newblock \bibinfo{title}{Line structures in recurrence plots},
\newblock \bibinfo{journal}{Physics Letters A} \bibinfo{volume}{336} (\bibinfo{year}{2005}) \bibinfo{pages}{349--357}. \DOIprefix\doi{10.1016/j.physleta.2004.12.056}.
\bibitem[{Gottwald and Melbourne(2004)}]{gottwald2004}
\bibinfo{author}{G.~A. Gottwald}, \bibinfo{author}{I.~Melbourne},
\newblock \bibinfo{title}{A new test for chaos in deterministic systems},
\newblock \bibinfo{journal}{Proc. R. Soc. Lond. Ser. Math. Phys. Eng. Sci.} \bibinfo{volume}{460} (\bibinfo{year}{2004}) \bibinfo{pages}{603--611}. \DOIprefix\doi{10.1098/rspa.2003.1183}.
\bibitem[{Majda and Lee(2014)}]{majda2014}
\bibinfo{author}{A.~J. Majda}, \bibinfo{author}{Y.~Lee},
\newblock \bibinfo{title}{Conceptual dynamical models for turbulence},
\newblock \bibinfo{journal}{Proc. Natl. Acad. Sci.} \bibinfo{volume}{111} (\bibinfo{year}{2014}) \bibinfo{pages}{6548--6553}. \DOIprefix\doi{10.1073/pnas.1404914111}.

\end{thebibliography}

\clearpage
\appendix
\renewcommand\thesection{\arabic{section}}
\setcounter{section}{0} 
\crefalias{section}{appendix}%
\crefalias{subsection}{appendix}

\begin{center}
\section*{Supplementary Information}    
\end{center}

\section{Reduced-order model}
\label{supp:doc1}

Here we describe the ROM for  a system in modal form with convective transport and impulse-driven forcing. We first initialize system parameters, then evolve convective elements, compute coupling impulses, and update the modal amplitudes by time integration.
The model is motivated from bluff-body combustor dynamics using $N$ acoustic modes coupled to convected vortices. The coupling occurs via heat-release impulses when vortices impinge at a fixed location~$L_c$.

\subsection*{1. Initialize System Parameters}
Given
\[
\gamma,\; c_0,\; L,\; L_c,\; d,\; \xi_1,\; St,\; N,\; \beta,\; \rho_0,\; U_0,\; \alpha_0,\; \sigma_\alpha,
\]
set the reference pressure
\[
p_0=\frac{\rho_0 c_0^2}{\gamma}, \qquad
c=\frac{2(\gamma-1)\beta}{L\,p_0}\quad \text{(impulse coefficient)}.
\]

\subsection*{2. Modal Properties}
For modes $n=1,\dots,N$:
\[
k_n=\frac{(2n-1)\pi}{2L}, \qquad \omega_n=c_0 k_n.
\]

\subsection*{3. Initial Conditions}
\[
g_1=\varepsilon_0,\quad g_{n>1}=0, \quad \dot g_n=0 \;\; \forall n.
\]

\subsection*{4. Time Marching (for $t=0:dt:t_{\text{end}}$)}
\paragraph{(a) Fields at position $x$}
\[
\phi_1(x,t)=p_0 \sum_{n=1}^N \frac{\dot g_n \cos(k_n x)}{\omega_n}, \qquad
\phi_2(x,t)=-\frac{c_0}{\gamma}\sum_{n=1}^N g_n \sin(k_n x).
\]

\paragraph{(b) Convective element generation at $x=0$:}
\[
u_{\text{local}}=U_0+\phi_2(0,t), \quad
\Gamma(t)=\Gamma(t-dt)+\kappa\,u_{\text{local}}^2\,dt, \quad
\Gamma_{\text{crit}}=\frac{u_{\text{local}}\, d}{2\,St}.
\]

\paragraph{(c) Element trajectories} For each element $(x_i,\Gamma_i)$,
\[
\alpha_i=\alpha_0+\sigma_\alpha \,\mathcal{N}(0,1), \qquad
\dot x_i=\alpha_i U_0+\phi_2(x_i,t).
\]

\paragraph{(d) Element–mode coupling at $x_i=L_c$}
\[
\psi_n(L_c)=\cos(k_n L_c), \qquad
I_n=c\,\Gamma_i\,\omega_n\,\psi_n(L_c), \qquad
\dot g_n \leftarrow \dot g_n + I_n.
\]

\paragraph{(e) Modal evolution}
\[
\xi_n=\xi_1\, f_{\text{damp}}(n), \quad f_{\text{damp}}(n)=(2n-1)^2 \;\; \text{(typ.)}
\]
\[
\ddot g_n=-\xi_n \dot g_n - \omega_n^2 g_n + F_n(t).
\]
Advance $(g_n,\dot g_n)$ with RK4.

\subsection*{Reference Parameters}
Reference parameter set (bluff-body combustor inspired) are provided below.
\begin{table}[h]
\centering
\label{tab:reference_params}
\begin{tabular}{@{}lll@{}}
\toprule
\textbf{Parameter} & \textbf{Value} & \textbf{Description} \\
\midrule
$\gamma$ & 1.4 & Specific heat ratio (air) \\
$c_0$ & 700.0 m/s & Sound speed \\
$L$ & 0.7 m & Domain length \\
$L_c$ & 0.05 m & Coupling location \\
$d$ & 0.025 m & Characteristic length \\
$\xi_1$ & 29.0 s$^{-1}$ & Base damping \\
$St$ & 0.35 & Strouhal number \\
$N$ & 10 & Number of modes \\
$\beta$ & $6\times 10^3$ & Coupling (heat-release) coeff. \\
$\rho_0$ & 1.225 kg/m$^3$ & Reference density \\
$U_0$ & 8.0 m/s & Mean convective speed \\
$\alpha_0$ & 0.2 & Mean convection ratio \\
$\sigma_\alpha$ & 0.02 & Turbulence intensity \\
$\varepsilon_0$ & 0.001 & Initial perturbation \\
$\kappa$ & 0.5 & Generation coefficient \\
\bottomrule
\end{tabular}
\end{table}

\clearpage

\section{CNN Architecture}
\label{supp:cnn}

We design a CNN to classify $450 \times 450$ recurrence matrices from dynamical systems into four dynamical regimes. The network employs hierarchical feature extraction via convolutional layers, batch normalization, and pooling, followed by a fully connected classification head. Regularization through dropout and batch normalization mitigates overfitting. For training we use the Adam optimizer with cross-entropy loss.

\subsection*{Input Representation}

The input tensor is:
\[
\mathbf{X} \in \mathbb{R}^{1 \times 450 \times 450}
\]
where $X_{ij}$ denotes the recurrence between states at times $i$ and $j$. All matrices are normalized to floating-point tensors prior to training.

\subsection*{Convolutional Feature Extraction}

The network uses three convolutional blocks. Each block applies:
\[
\mathbf{Z}^{(\ell)} = \text{MaxPool}\left(\text{ReLU}\left(\text{BN}\left(\mathbf{W}^{(\ell)} * \mathbf{H}^{(\ell-1)} + \mathbf{b}^{(\ell)}\right)\right)\right)
\]
where $*$ denotes 2D convolution.

\noindent\textbf{Layer specifications:}
\begin{itemize}
    \item {Block 1:} $\text{Conv2d}(1, 32, k=3, p=1) \to \text{BN} \to \text{ReLU} \to \text{MaxPool}(k=4,s=4)$; output $32 \times 112 \times 112$.
    \item {Block 2:} $\text{Conv2d}(32, 64, k=3, p=1) \to \text{BN} \to \text{ReLU} \to \text{MaxPool}(k=2,s=2)$; output $64 \times 56 \times 56$.
    \item {Block 3:} $\text{Conv2d}(64, 128, k=3, p=1) \to \text{BN} \to \text{ReLU} \to \text{MaxPool}(k=2,s=2)$; output $128 \times 28 \times 28$.
\end{itemize}

\subsection*{Classification Head}

The output of Block 3 is flattened to:
\[
\mathbf{f} \in \mathbb{R}^{100352}, \quad 100352 = 128 \times 28 \times 28
\]
The classifier applies:
\[
\mathbf{h} = \text{ReLU}(\text{BN}(\mathbf{W}_1 \mathbf{f} + \mathbf{b}_1)), \quad
\mathbf{h}_{\text{drop}} = \text{Dropout}(\mathbf{h}; p=0.5), \quad
\mathbf{y} = \mathbf{W}_2 \mathbf{h}_{\text{drop}} + \mathbf{b}_2
\]
where $\mathbf{y} \in \mathbb{R}^4$ are class logits.

\subsection*{Training}

We use cross-entropy loss:
\[
\mathcal{L}(\mathbf{y}, t) = -\log\left(\frac{\exp(y_t)}{\sum_{j=1}^4 \exp(y_j)}\right)
\]
with Adam optimization:
\[
\theta_{t+1} = \theta_t - \frac{\alpha}{\sqrt{\hat{v}_t} + \epsilon} \hat{m}_t
\]
and parameters listed below 
\begin{table}[h]
\centering
\label{tab:cnnparams}
\begin{tabular}{@{}ll@{}}
\toprule
\textbf{Parameter} & \textbf{Value} \\ 
\midrule
Input size & $1 \times 450 \times 450$ \\
Conv block channels & (32, 64, 128) \\
Kernel size & $3 \times 3$ \\
Pooling factors & (4, 2, 2) \\
Activation & ReLU \\
Batch normalization & After each conv and FC1 \\
Dropout rate & 0.5 (before final FC) \\
FC layer sizes & 100352 $\to$ 128 $\to$ 4 \\
Loss function & Cross-entropy \\
Optimizer & Adam \\
Learning rate $\alpha$ & 0.001 \\
Adam $\beta_1, \beta_2$ & 0.9, 0.999 \\
Adam $\epsilon$ & $10^{-8}$ \\
Batch size & 32 \\
\bottomrule
\end{tabular}
\end{table}

\clearpage
\section{TRIAD model}
\label{supp:doc2}

We simulate a stochastic complex triad system. Three nonlinearly coupled oscillators subject to deterministic forcing and stochastic perturbations. The model exhibits periodic oscillations, chaotic fluctuations, and intermittent bursts. A semi-implicit Euler scheme is used to integrate the system, treating linear terms implicitly and nonlinear terms explicitly.

\subsection*{Mathematical Formulation}

Each oscillator state $u_j(t)$ ($j=1,2,3$) is complex variable satisfying:
\begin{align}
\frac{d u_1}{dt} &= (-\gamma_1 + i\omega_1)u_1 + L_{12}u_2 + L_{13}u_3 + Iu_1u_2 + F_1 + \sigma_1\xi_1(t), \\
\frac{d u_2}{dt} &= (-\gamma_2 + i\omega_2)u_2 - L_{12}u_1 + L_{23}u_3 - Iu_1^2 + \sigma_2\xi_2(t), \\
\frac{d u_3}{dt} &= (-\gamma_3 + i\omega_3)u_3 - L_{13}u_1 - L_{23}u_2 + \sigma_3\xi_3(t),
\end{align}
where $\xi_j(t) = \eta_{2j-1}(t) + i\eta_{2j}(t)$, $\eta_k \sim \mathcal{N}(0,1)$ are independent Gaussian processes.

\subsection*{Numerical Integration}

The semi-implicit Euler update is:
\begin{align}
u_1^{n+1} &= \frac{u_1^n + \Delta t(L_{12}u_2^n + L_{13}u_3^n + Iu_1^n u_2^n + F_1) + \sigma_1\Delta W_1^n}{1 + (\gamma_1 - i\omega_1)\Delta t}, \\
u_2^{n+1} &= \frac{u_2^n + \Delta t(-L_{12}u_1^n + L_{23}u_3^n - I(u_1^n)^2) + \sigma_2\Delta W_2^n}{1 + (\gamma_2 - i\omega_2)\Delta t}, \\
u_3^{n+1} &= \frac{u_3^n + \Delta t(-L_{13}u_1^n - L_{23}u_2^n) + \sigma_3\Delta W_3^n}{1 + (\gamma_3 - i\omega_3)\Delta t},
\end{align}
with $\Delta W_j^n = \sqrt{\Delta t}\,\xi_j^n$.
For each denominator $D_j = 1 + (\gamma_j - i\omega_j)\Delta t$, we note the following that is used for stability analysis:
\[
|D_j| = \sqrt{(1+\gamma_j\Delta t)^2 + (\omega_j\Delta t)^2}, \quad
\theta_j = \arctan\!\left(\frac{\omega_j\Delta t}{1+\gamma_j\Delta t}\right).
\]

\subsection*{Parameters}

The model parameter are provided below, which produce periodic background oscillations with intermittent bursts. 
\begin{table}[h]
\centering
\begin{tabular}{@{}lll@{}}
\toprule
\textbf{Parameter} & \textbf{Value} & \textbf{Description} \\ 
\midrule
$\gamma_1, \gamma_2, \gamma_3$ & 0.05, 0.08, 0.03 & Damping coefficients \\
$\omega_1, \omega_2, \omega_3$ & 1.0, 0.5, 1.5 & Natural frequencies \\
$L_{12}, L_{13}, L_{23}$       & 0.25, 0.15, 0.12 & Linear coupling \\
$I$                            & 0.6              & Nonlinear coupling  \\
$F_1$                          & 0.3              & External forcing  \\
$\sigma_1, \sigma_2, \sigma_3$ & 0.15, 0.12, 0.10 & Noise strength \\
$\Delta t$                     & 0.001            & Time step \\ 
\bottomrule
\end{tabular}
\end{table}

\clearpage

\section{Anisotropic flow model}
\label{supp:doc3}

We consider a stochastic low-dimensional model consisting of a mean flow component $u_m$ coupled to $K$ turbulent modes $u_k$ ($k=1,\dots,K$). Nonlinear mode coupling, cubic damping of the mean flow, and stochastic forcing generate intermittent bursting. The system is integrated using the Euler--Maruyama method.

\subsection*{Governing Equations}
\begin{align}
\frac{du_m}{dt} &= -d_m u_m + \gamma \sum_{k=1}^K u_k^2 - \alpha_m u_m^3 + F_m + \sigma_m\,dW_m, \\
\frac{du_k}{dt} &= -d_k u_k - \gamma u_m u_k + \sigma_k\,dW_k, \quad k = 1,\dots,K,
\end{align}
where $d_m$ is the mean-flow damping, $d_k$ the mode damping, $\gamma$ the nonlinear coupling, $\alpha_m$ the cubic damping coefficient, $F_m$ the external forcing, $\sigma_m$ and $\sigma_k$ the noise intensities, and $W_m$, $W_k$ independent Wiener processes.

\paragraph{Mode damping}
\begin{equation}
d_k = 1 + 0.02 k^2
\end{equation}
This models increasing dissipation with mode number.

\paragraph{Noise scaling}
\begin{equation}
\sigma_k = \sqrt{\frac{d_k \cdot 0.004}{(1+k)^{5/3}}},
\end{equation}
This produces a $-5/3$ spectral slope (Kolmogorov scaling) in the inertial range.

\subsection*{Numerical Integration}
Euler--Maruyama update for time step $\Delta t$:
\begin{align}
u_m^{n+1} &= u_m^n + \Delta t\biggl(-d_m u_m^n + \gamma\sum_{k}(u_k^n)^2 - \alpha_m (u_m^n)^3 + F_m\biggr) + \sigma_m\sqrt{\Delta t}\,\xi_m^n, \\
u_k^{n+1} &= u_k^n + \Delta t\!\left(-d_k u_k^n - \gamma u_m^n u_k^n\right) + \sigma_k\sqrt{\Delta t}\,\xi_k^n,
\end{align}
where $\xi_m^n, \xi_k^n \sim \mathcal{N}(0,1)$ are independent at each step. Simulations use $\Delta t = 0.005$, sufficient for stability.

\subsection*{Parameters}
\begin{itemize}
    \item Number of turbulent modes: $K=5$, total system dimension $N=6$
    \item $F_m = -0.055$, $d_m = -0.1$, $\alpha_m = 0.05$, $\gamma = 1.5$
    \item Initial conditions: $u_m(0) = -\frac{2}{3}$, $u_k(0) = 0$
    \item Integration time: $T=1000$ units, $200{,}000$ steps; initial transients discarded
\end{itemize}

\clearpage

\section{Kolmogorov flow model}
\label{supp:doc4}

We simulate the 2D incompressible Kolmogorov flow on a doubly periodic domain using a pseudospectral method. The Navier-Stokes equations are solved in a periodic domain with sinusoidal forcing. The implementation involves a divergence-free projection to enforce incompressibility, $2/3$-rule antialiasing for the nonlinear terms, and time advancement via the fourth-order Runge--Kutta (RK4) scheme. The simulation tracks energy dissipation.

\subsection*{Governing Equations}
\begin{align}
\nabla \cdot \mathbf{u} &= 0, \\
\frac{\partial \mathbf{u}}{\partial t} + (\mathbf{u} \cdot \nabla)\mathbf{u} &= -\nabla p + \frac{1}{Re}\nabla^2 \mathbf{u} + \mathbf{F},
\end{align}
with $\mathbf{F} = (0, F_0 \sin(n x))$. Here $Re$ is the Reynolds number and $n$ the forcing wavenumber.

\subsection*{Spectral Representation}
On a doubly-periodic domain $[0, 2\pi] \times [0, 2\pi]$:
\[
f(x,y) = \sum_{k_x=-N}^{N} \sum_{k_y=-N}^{N} \hat{f}(k_x, k_y) e^{i(k_x x + k_y y)}.
\]
Spatial derivatives become:
\[
\widehat{\partial_x f} = i k_x \hat{f}, \quad
\widehat{\partial_y f} = i k_y \hat{f}, \quad
\widehat{\nabla^2 f} = -(k_x^2 + k_y^2) \hat{f}.
\]

\subsection*{Divergence-Free Projection}
To enforce incompressibility, a projection operator is applied in the spectral domain. For any vector field with  Fourier components $(\hat{f}_x,\hat{f}_y)$:
\begin{align}
P_1 &= \frac{k_y^2}{k_x^2 + k_y^2}, \quad
P_2 = -\frac{k_x k_y}{k_x^2 + k_y^2}, \quad
P_3 = \frac{k_x^2}{k_x^2 + k_y^2}, \\
\hat{u}_{\text{div-free}} &= P_1 \hat{f}_x + P_2 \hat{f}_y, \\
\hat{v}_{\text{div-free}} &= P_2 \hat{f}_x + P_3 \hat{f}_y,
\end{align}
with $(k_x,k_y)=(0,0)$ set to zero.

\subsection*{Energy dissipation}
The vorticity in spectral form:
\[
\hat{\omega} = i k_y \hat{u} - i k_x \hat{v}.
\]
We then compute the energy dissipation from:
\[
D(t) = \frac{\nu}{L^2} \iint |\omega|^2 \, d\mathbf{x}, \quad \nu = 1/Re.
\]

\subsection*{Initialization}
A random divergence-free field is generated in spectral space with amplitude
\[
A(k_x,k_y) = A_{\text{mag}} \cdot 4(2N+1)^2 \frac{1}{\sqrt{2\pi\sigma^2}} \exp\biggl(-\frac{(k_x-c_1)^2 + (k_y-c_2)^2}{2\sigma^2}\biggr) ,
\]
random phases $\phi_u,\phi_v \in [0,2\pi]$, and then the divergence free projection is applied to enforce incompressibility.

\subsection*{Parameters}
\begin{itemize}
    \item $Re = 500$, $N=12$ (resulting in a $25\times25$ grid)
    \item Forcing wavenumber $n=4$
    \item $T=1000$, $\Delta t = 0.01$
    \item Initial field: $\sigma=1.0$, $A_{\text{mag}}=0.1$, $(c_1,c_2)=(0,3)$
\end{itemize}

\end{document}